# Study of GeSn Selective Area Growth with Demonstration of SWIR Light Detection

*Hryhorii Stanchu [1,‡], Quang Minh Thai [2,‡], Rajesh Kumar [2], Fernando M. de Oliveira [1], Kushal Dahal [2,3], Xuehuan Ma [2], Sudip Acharya [2,3], Justin Rudie [2,3], Alexander Golden [2,3], Joshua M Grant [2], Matthew Cook [4], Stephen Margiotta [4], Xiaoxin Wang [5], Jifeng Liu [5], Perry C. Grant [6], Baohua Li [6], Wei Du [1,2,3], Gregory Salamo [1,3,7] and Shui-Qing Yu [1,2,3*]*

[1]Arkansas Materials Institute, University of Arkansas, Fayetteville, Arkansas 72701, USA

[2]Department of Electrical Engineering and Computer Science, University of Arkansas, Fayetteville, Arkansas 72701, USA

[3]Material Science and Engineering Program, University of Arkansas, Fayetteville, Arkansas 72701, USA

[4] Lincoln Laboratory, Massachusetts Institute of Technology, 244 Wood Street, Lexington, Massachusetts 02421, USA

[5] Thayer School of Engineering, Dartmouth College, Hanover, New Hampshire 03755, USA

[6] Arktonics, LLC, 1339 South Pinnacle Dr., Fayetteville, Arkansas 72701, USA

[7] Department of Physics, University of Arkansas, Fayetteville, Arkansas 72701, USA

**ABSTRACT.** As germanium-tin (GeSn) epitaxial growth quality continuously improves, the search for an efficient integration strategy of GeSn optoelectronics devices into complementary metal-oxide-semiconductor (CMOS) manufacturing line also accelerates. Selective area growth (SAG) on patterned substrate emerges as a promising approach for this quest, with locally controlled growth of GeSn laser/ detector suitable for either co-integration with silicon-based waveguide structure or stand-alone module like focal plane array. In this work, we report successful GeSn SAG with Sn content ranging from 3.2% to 8.7% of good optical quality, with demonstration of tunable GeSn SAG photoluminescence and GeSn SAG photoconductor device, the latter with detection cutoff wavelength up to 2 μm of wavelength. In addition, we present a comprehensive study of GeSn SAG condition at different window sizes – from 2 μm to 100 μm – and shapes – circle, square, octagon and rectangle. Presence of loading effect is revealed, where GeSn growth rate increases as pattern fill factor and window size shrink. It introduces a different growth condition compared to thin film growth, which can weaken or inhibit Sn incorporation at very small window size and induce Sn segregation in high Sn content SAG growth.

With tunable band gap in the infrared range, the ability to transform into direct-gap material at high tin (Sn) incorporation and in particular the capability of monolithic growth on silicon (Si) substrate, germanium-tin (GeSn) alloy is a promising candidate for the development of CMOS compatible infrared laser and detector. Important milestones achieved over the last ten years helped to further consolidate its position as the forefront material for such objective. GeSn lasers advanced from the first demonstration of a GeSn optically pumped laser at cryogenic temperature in 2015 [1] to room-temperature optically pumped GeSn lasers [2–4] and electrically injected GeSn lasers [5–7]. At the same time, GeSn detector performance showed steady progress [8–10], reducing the performance gap with III-V and II-VI photodetector in short-wave infrared (SWIR) and extended SWIR (e-SWIR) range.

As the GeSn epitaxial growth quality and device performance continue to improve, the other front of the quest - developing integration strategy for GeSn optoelectronics devices to be fully adopted in CMOS manufacturing line – slowly comes under spotlight and attracts increasing research interests. Selective area growth (SAG) on patterned substrate, as previously demonstrated for direct growth of III-V devices on Si [11–17], emerges as a suitable strategy for this task. Here, GeSn is locally grown on patterned Si substrate with micro or nano-size window openings, which can be designed and adapted for either laser/detector integration in photonics integrated circuit (PIC) or for stand-alone module like focal plane array (FPA) imaging system [14]. A previous attempt on GeSn nano-sized SAG has been reported from our group, showing unsuccessful Sn incorporation when using standard GeSn thin film growth recipe, however [18]. It suggested substantially different growth characteristics between GeSn thin film growth and SAG, which can be strongly correlated to the shape and size of the oxide window opening.

Understanding how they evolve from one to another is thus the key for successful micro- and eventually nano-patterned GeSn SAG.

In this paper, we reported successful GeSn SAG with Sn content up to 8.7% of good optical quality, demonstrated through the presence of tunable photoluminescence spectra and GeSn SAG photoconductor operation with detection range up to 2 μm of wavelength. In addition, a comprehensive study on growth condition under various window shapes – circle, octagon, square and rectangle - and window sizes from 2 μm to 100 μm was also presented. We observed the presence of a loading effect, where the GeSn growth rate noticeably increased when window size/ window shape fill factor shrank, resulting in either a reduction/absence of Sn incorporation for SAG in very small window opening, or an appearance of Sn droplets in high Sn content growth recipe.

## RESULTS AND DISCUSSION

### Material growth – Average XRD characteristics of GeSn SAG samples

To calibrate the growth recipe prior to SAG, GeSn thin film samples were first grown with Sn content ranging from 2.9 % Sn to 12.2 % Sn (**Figure 1**), with detailed growth parameters have been reported elsewhere [19]. Here, GeSn thin film layer thicknesses were between 21 nm to 48 nm, below the critical thickness for strain relaxation to avoid spontaneous relaxation enhanced (SRE) effect and gradual distribution of Sn content [19–21], therefore maintaining a constant Sn content in the grown layer for precise calibration.

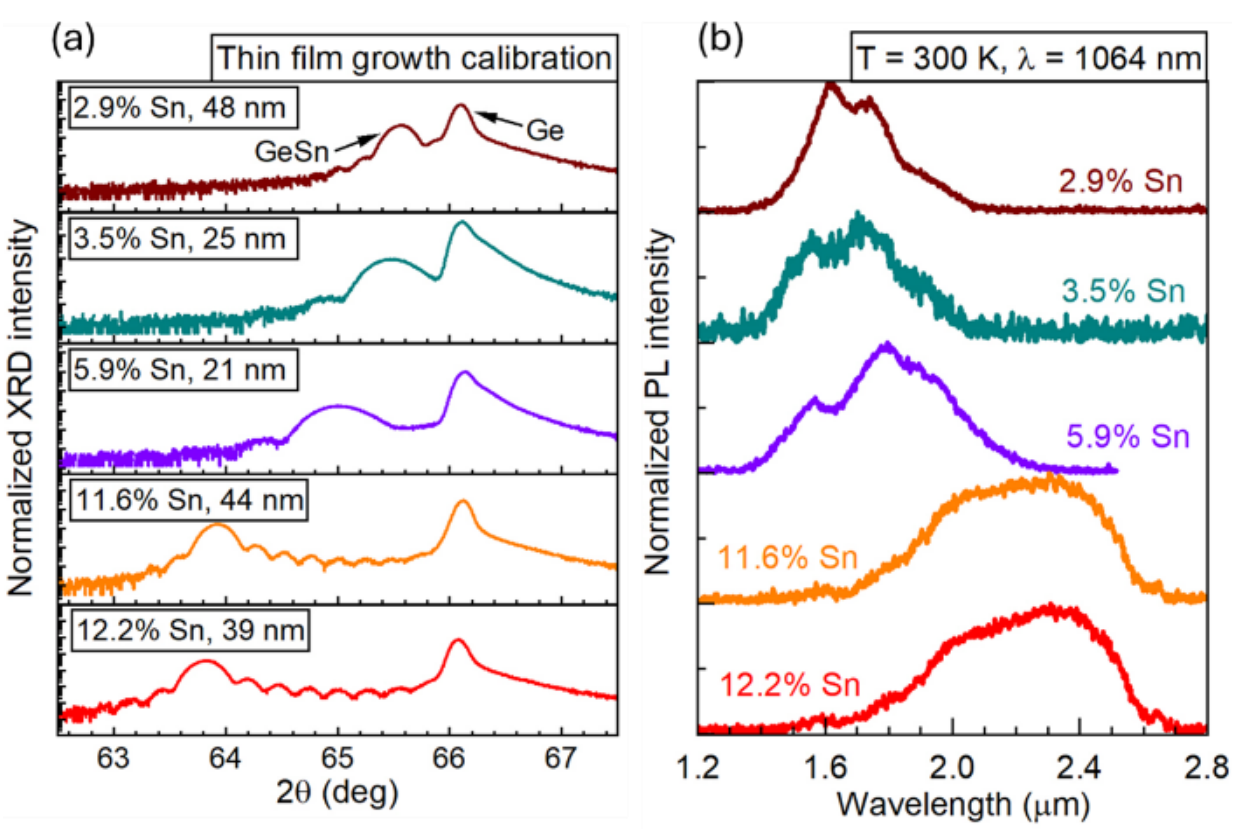

***Figure 1.*** *(a) ω-2θ XRD scans and (b) PL spectra of thin film GeSn samples from 2.9% Sn to 12.2% Sn, for the growth recipe calibration prior to GeSn SAG.*

To investigate the influence of window shape and size on GeSn SAG characteristics, lithography mask for $SiO_2$ window etching shape were designed as circle, octagon, square, and rectangle, with sizes ranging from 2 µm to 100 µm (**Figure 2**). Based on the thin film calibration results, GeSn SAG were then conducted on Si wafer with patterned $SiO_2$ window in a commercial ASM Epsilon 2000 reduced pressure chemical vapor deposition (RPCVD) reactor, using growth recipe targeting 2%, 4%, 5%, 8% and 10% nominal Sn content with similar 2-steps growth of Ge buffer/GeSn, alongside a Ge SAG sample as reference. $GeH_4$ and $SnCl_4$ were used as Ge and Sn gas precursors, respectively. HCl flow was introduced during the growth process to maintain selectivity, inhibiting Ge, GeSn, and Sn growth on $SiO_2$ surface. $SiO_2$ layer thickness was measured to be 570 nm.  Growth time for SAG samples was calculated to target a confined growth of GeSn layer inside the window opening, hypothesizing a similar growth rate between GeSn thin film and GeSn SAG.

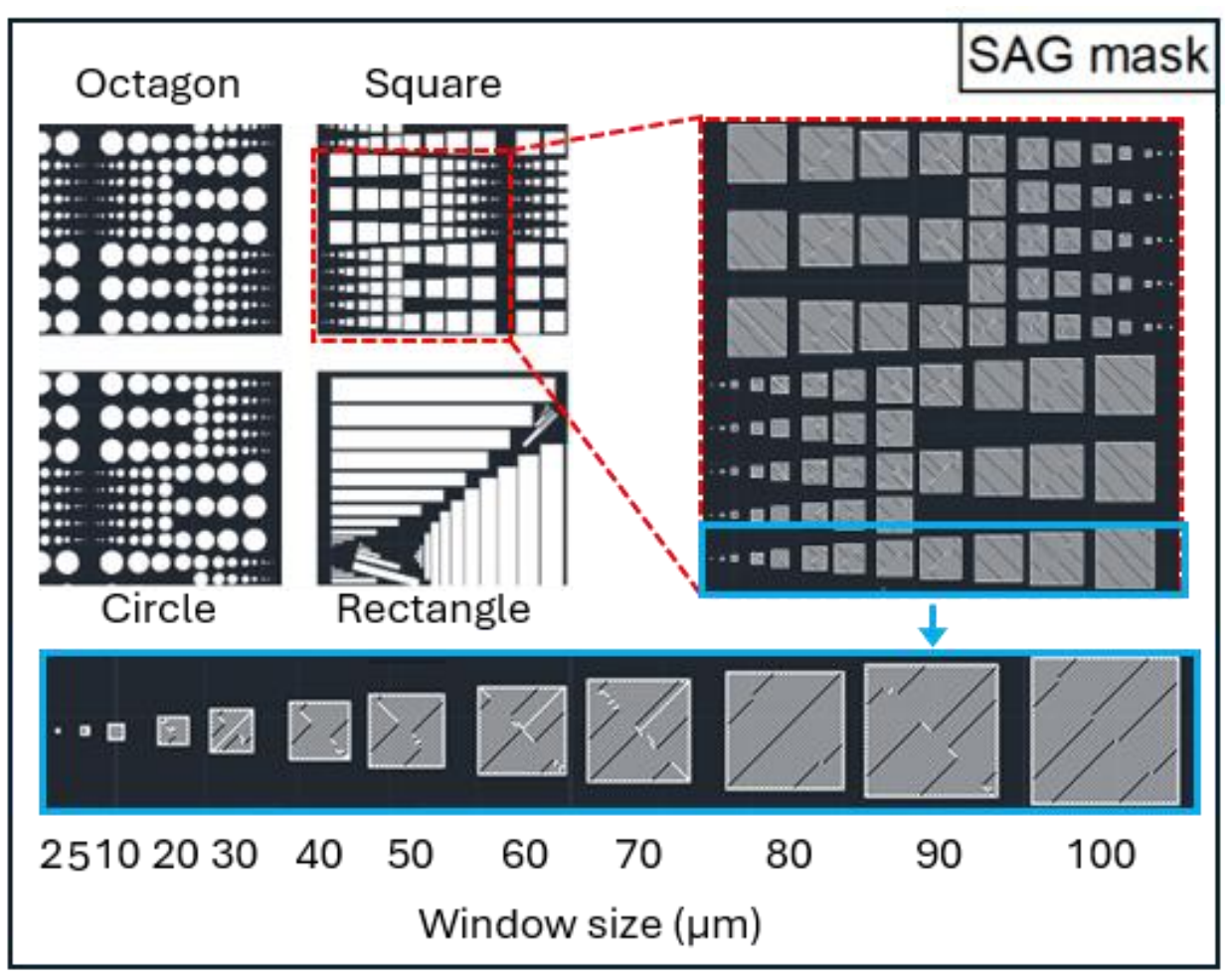

***Figure 2.*** *Mask design for $SiO_2$ window etching in Ge and GeSn SAG samples, showing octagon, square, circle, and rectangle patterns of sizes varying from 2 µm to 100 µm.*

X-ray diffraction (XRD) was then performed on all SAG samples: Sn contents of SAG samples - from 3.2% to 8.7% - were extracted from (224) reciprocal space mapping (RSM) data and shown in **Figures 3a-e**. Here, XRD signal represented an average response from many SAG windows of different sizes, due to large X-ray beam size of approximately 5 $mm^2$. ω-2θ plots for selected window shapes were shown in **Figure 3f**, with 2θ peak positions on all window shapes and all SAG samples shown in **Figure 3g**: at low Sn content, below 6.7%, nearly constant 2θ peak position across all window shapes indicated that the morphology of SAG island had only negligeable effect on the Sn content and residual strain. At highest Sn content (sample 5 with 8.7% Sn), the 2θ peak shifted toward higher diffraction angles for circle and octagon windows, indicating a Sn depletion effect on the bulk of GeSn layer and thus a Sn segregation for these geometries. One can observe that SAG experimental Sn content closely followed the nominal target (**Figure 3h**), suggesting that GeSn growth recipe calibrated from thin film data remained valid for SAG at least at large size windows, as the average XRD response was strongly skewed towards large size SAG island with more material volume. Experimental Sn contents and nominal targets were provided in **Table 1**.

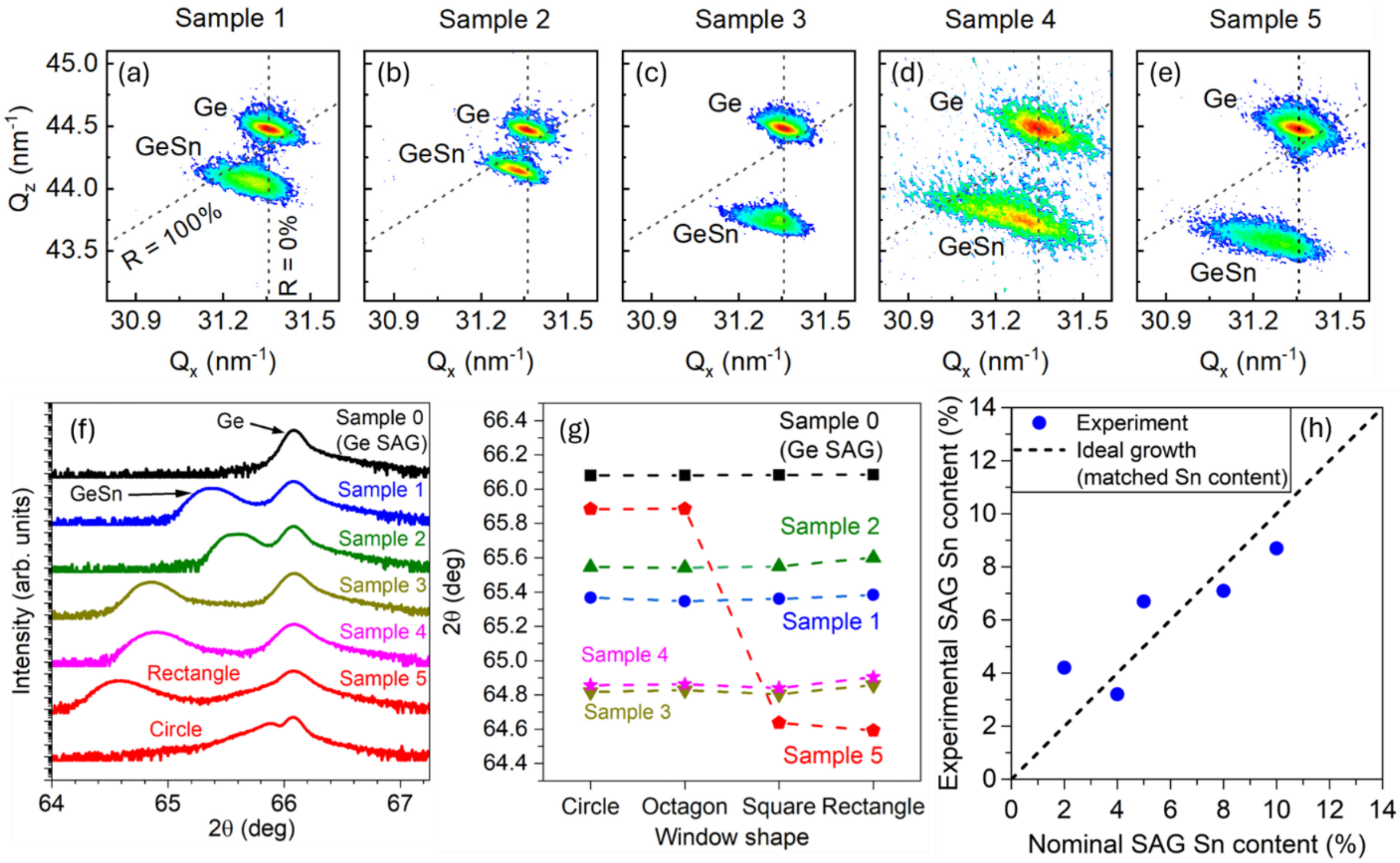

***Figure 3.*** *(a)-(e) (224) RSM plots of GeSn SAG samples, on selected window shapes: rectangle, octagon, rectangle, circle, rectangle for samples 1 to 5, respectively. (f) ω-2θ scans of GeSn samples, with Ge SAG sample as reference (recipe 0), for selected window shapes. (g) GeSn SAG peak positions evaluated from the ω-2θ profile, across all window shapes. In sample 5, clear shift of peak position to higher 2θ values indicated a reduction of Sn content in the GeSn layer - signature of Sn segregation. (h) Plot of GeSn SAG experimental Sn content extracted from (224) RSM plots, as a function of nominal target. Dashed lines showed ideal growth configuration where experimental Sn content matched nominal target.*

***Table 1*** *Summary of Sn content in GeSn SAG samples 1 to 5 in this work, compared to the nominal target. Sample 0 was Ge SAG.*

| SAG Sample | Nominal Sn content (%) | Experimental Sn content (%) |
|---|---|---|
| 0 | 0 (Ge) | |
| 1 | 2 | 4.2 |
| 2 | 4 | 3.2 |
| 3 | 5 | 6.7 |
| 4 | 8 | 7.1 |
| 5 | 10 | 8.7 |

**GeSn SAG photoluminescence and photoconductor characterization**

Photoluminescence (PL) spectra of SAG samples showed an expected red-shift of PL peak wavelength position at higher Sn content (**Figure 4**), increasing from 1.6 µm for Ge SAG (sample 0) to 2.1 µm for 8.7% Sn GeSn SAG (sample 5). It is interesting to note that SAG PL spectra, even at very low or no Sn content, were mainly composed of the Γ direct band gap emission peak, instead of both direct and L indirect band gap emission peaks as in bulk Ge reference sample. Considering the low radiative efficiency of the indirect transition, the reduced intensity of the indirect band-gap emission can be attributed to the small absorption volume, similar to what was previously observed on Ge SAG [22]. With increased Sn content, the PL contribution from the Ge buffer was significantly reduced, leaving only contribution from the GeSn SAG layer on the spectrum. It should be noted that due to the large laser spot size in our PL setup, the PL spectra shown here are the average response from SAG islands of different sizes, similar to prior XRD characterization.

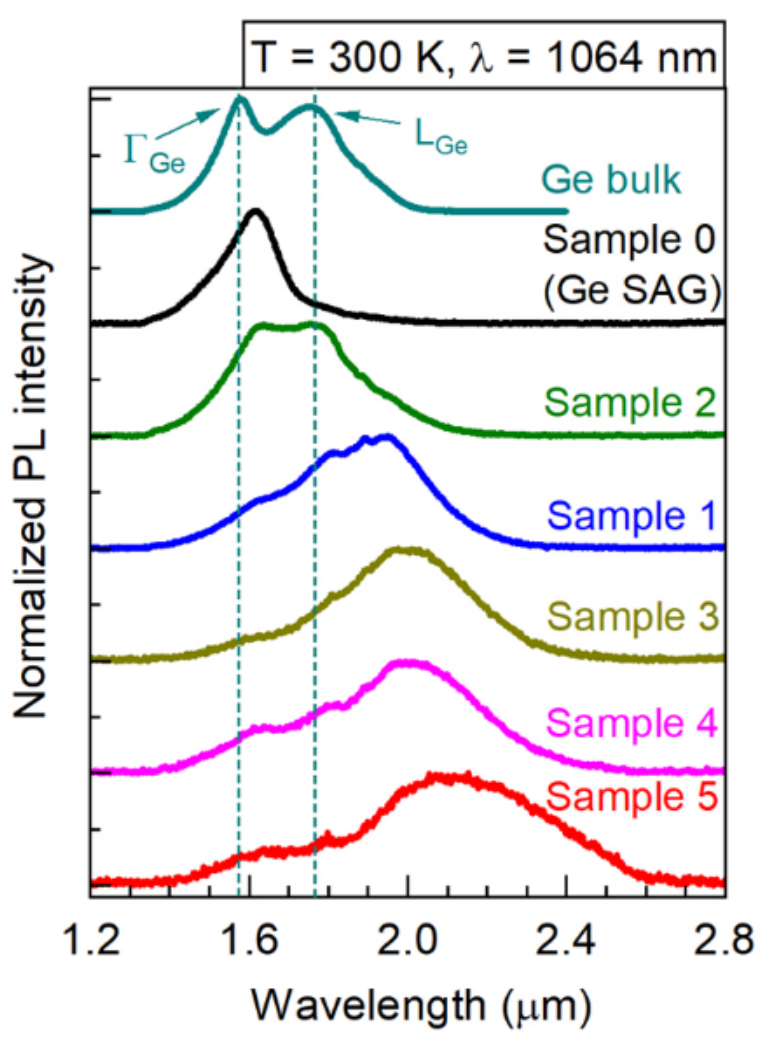

***Figure 4**. PL spectra of bulk Ge reference sample, Ge SAG and GeSn SAG samples at 300 K, measured under 1064 nm laser excitation. Dashed lines indicated Γ direct band gap and L indirect band gap emission peak position in bulk Ge reference sample.*

To further explore GeSn SAG quality for optoelectronics device application, we fabricated and characterized photoconductors fabricated from sample 1 (4.2% Sn) and sample 4 (7.1% Sn), with interdigitated electrode on square SAG islands of 100 µm size. Photoconductor characterization results for sample 1 (4.2 % Sn) were shown in **Figure 5**. The current-voltage (I-V) characteristics showed a clear difference between dark and light conditions, with device dark resistance varied from 738 kΩ to 53 kΩ between 77 K and 300 K, comparable to previous work on GeSn SWIR photoconductor [23] (**Figure 5a**). Responsivity up to 0.9 $A.W^{-1}$ was recorded at 300 K and shown in **Figure 5b**, with early saturation of responsivity likely due to minority carrier sweep out combined with Joule heating effect [8,24]. From FTIR spectral response plotted in **Figure 5c**, the detection cutoff wavelength - i.e. where FTIR relative intensity drops to 50% - was found to be 1.89 µm at 300 K, in line with the Sn content and the PL spectrum range shown earlier in **Figure 4**.

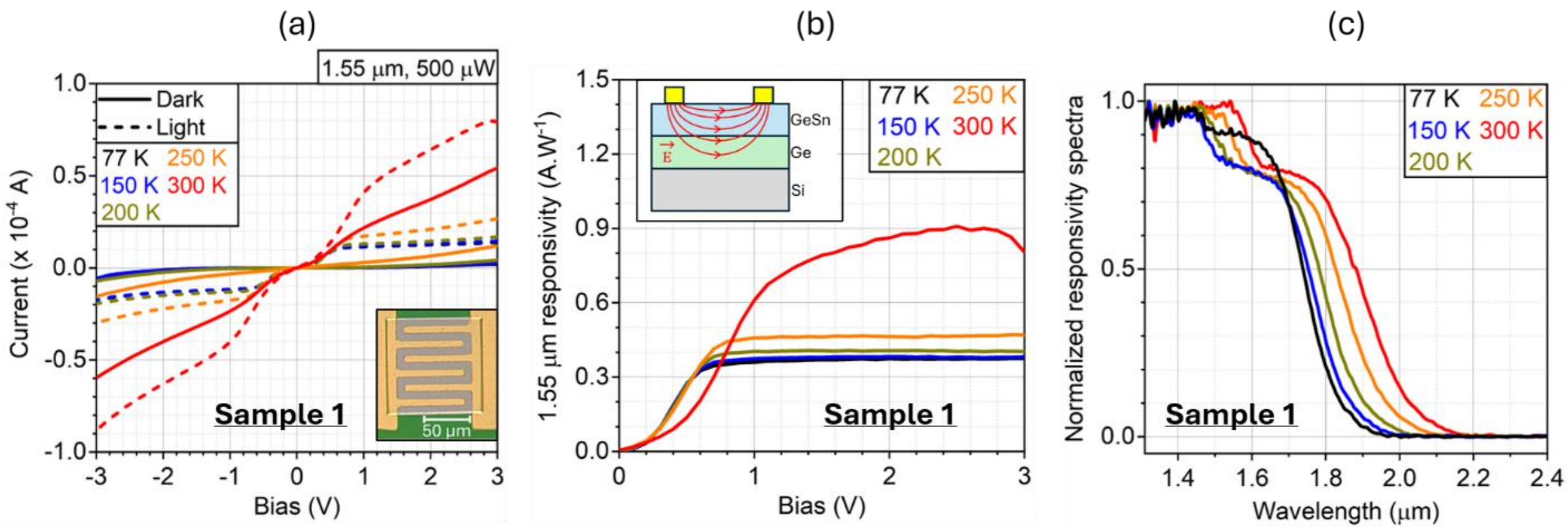

***Figure 5**. (a) Dark (solid) and light (dashed) current-voltage (I-V) characteristics from 77 K to 300 K of an interdigitated photoconductor (100 µm square – microscopy image of the device was shown in the inset) fabricated from GeSn SAG sample 1 (4.2% Sn). Light I-V was taken under 1.55 µm illumination with an optical power of 500 µW. (b) 1.55 µm responsivity as function of bias for sample 1 photoconductor, from 77 K to 300 K. Ratio of exposed device area over the laser beam size (320 µm diameter) has been taken into the calculation of responsivity. A schematic of electric field contour lines in GeSn SAG structure was provided in the inset (c) Normalized FTIR responsivity spectra of sample 1 photoconductor from 77 K to 300 K.*

It is interesting to note that the responsivity at 1.55 µm wavelength increased with temperature for sample 1 photoconductor. Here, as the grown GeSn layer was relatively thin, and as Ge and GeSn layer thickness were comparable, we expected that the electric field contour and thus current flow path penetrated both the GeSn layer and Ge buffer (**Figure 5b - inset**): as a result, both layers contributed to the responsivity at 1.55 µm wavelength. At cryogenic temperature, Ge detection cutoff wavelength was below 1.55 µm [25]; optical absorption solely took place in the GeSn layer and thus explained a weaker responsivity. As temperature increases,

Ge detection cutoff wavelength increased and thus contributed more to the responsivity at 1.55 µm wavelength, explaining the increase in responsivity.

Characterization results for sample 4 (7.1% Sn) photoconductor were shown in **Figure 6**. A noticeable increase of dark current compared to sample 1, with appearance of Schottky contact behavior at high bias (beyond 1 V) was observed (**Figure 6a**). It might hint at higher background p-doping concentration from a higher defect density in sample 4, which lowered the sample dark resistance and simultaneously introduced additional surface trap state, the latter can lead to formation of Schottky contact due to Fermi level pinning. Responsivity at 1.55 µm wavelength of sample 4 showed similar temperature-dependent behavior as sample 1, due to contribution from both Ge and GeSn layer (**Figure 6b**). At 2 µm wavelength, a responsivity up to 0.2 $A.W^{-1}$ was recorded a 300 K and decreased at lower temperatures (**Figure 6c**). It was explained from the responsivity spectra of sample 4 (**Figure 6d**): higher detection cutoff wavelength at 2 µm at 300 K was observed compared to sample 1, due to higher Sn content. As temperature decreased, GeSn band gap increased and lead to a lower detection cutoff wavelength below 2 µm, explaining the temperature-dependent behavior of 2 µm responsivity. Finally, we calculated the specific detectivity $D^*$ of both sample 1 and sample 4 photoconductors using the following formula [26]:

$$D^* = \frac{\mathcal{R}\sqrt{A}}{\sqrt{2q|I_d| + \frac{4k_BT}{R_d}}} \quad (1)$$

where $\mathcal{R}$, $A$, $|I_d|$, $R_d$, $T$ were the device responsivity, area, dark current (absolute value), dark resistance and temperature. For sample 4, the dark resistance was calculated from the linear part of IV characteristics between -1 V and 1 V. $D^*$ spectra at 300 K for sample 1 and sample 4 were calculated at 1 V bias and 0.8 V bias respectively and were plotted in **Figure 6e**: comparable performance to PbSe photoconductor in the SWIR and early extended SWIR range was observed, alongside a better performance than previous results for GeSn photoconductor reported from our group [8].

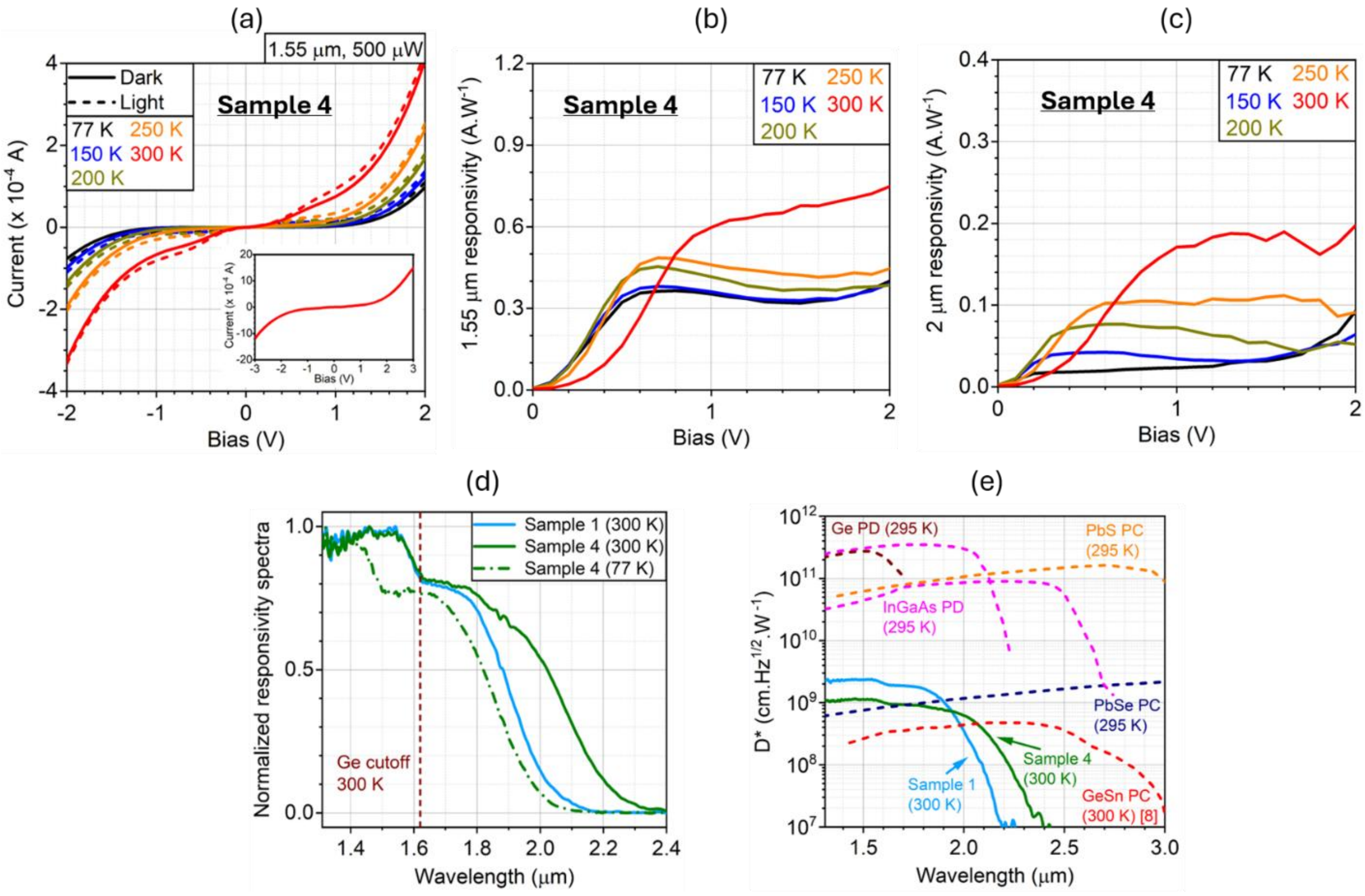

***Figure 6**. (a) Dark (solid) and light (dashed) current-voltage (I-V) characteristics from 77 K to 300 K of an interdigitated photoconductor (100 μm square) fabricated from GeSn SAG sample 4 (7.1% Sn). Full dark I-V characteristics from -3 V to 3 V at 300 K was provided in the inset. (b) 1.55 μm responsivity as function of bias for sample 4 photoconductor, from 77 K to 300 K. (c) 2 μm responsivity as function of bias for sample 4 photoconductor, from 77 K to 300 K. (d) Comparison of normalized FTIR responsivity spectra between sample 1 (light blue) sample 4 (green) photoconductor at 300 K (solid lines). Green dashed-dot lines showed FTIR responsivity spectra of sample 4 at 77 K. The dashed line showed the detection cutoff wavelength of an in-house CVD-grown Ge photodiode detector (1.62 μm), for comparison purpose. Please note that*

*for sample 4, we observed a consistently weaker 2 μm responsivity when directly measured with the tunable laser, compared to the results extracted from FTIR spectra (around 24% weaker at 300 K, down to 2% weaker at 77 K). (e) D* spectra for sample 1 (light blue) and sample 4 (green) photoconductor, compared to previous GeSn photoconductor [8] and other commercial infrared photodiode and photoconductor. D* is calculated at 1 V bias for sample 1 and 0.8 V bias for sample 4.*

.

**Impact of window size and shape on GeSn SAG material properties**

To better understand the effect of window shape and size on Sn incorporation in GeSn SAG, which remained missing from earlier XRD data and PL/ photoconductor characterization, we used a combination of scanning electron microscopy (SEM), micro-Raman spectroscopy and transmission electron microscopy (TEM) characterization techniques. SEM images of lowest Sn content SAG sample (sample 2 of 3.2% Sn) and highest Sn content SAG sample (sample 5 of 8.7% Sn) were shown in **Figure 7**, on all window shapes with sizes between 2 μm and 30 μm. On sample 2, as the window size reduced, one can observe clear {111} facets on the grown SAG island, forming a "pyramid" or "roof" morphology. These are facets with lowest energy surfaces for FCC crystal [27], marking a deviation from the multi-faceted, "dome" shape previously observed on nano-patterned Ge SAG [14,22] and revealing a different strain and surface tension relaxation mechanism in GeSn SAG. It can be due to stronger lattice mismatch in GeSn/Ge on Si structure compared to Ge on Si structure. Energy dispersive X-ray spectroscopy (EDX) was performed on square windows of sample 2, showing successful Sn incorporation down to 10 μm of window size.

On sample 5, the growth characteristics was noticeably different. Sn segregation/droplets were observed on circle and octagon windows, as suggested earlier from XRD data shown in previous section. Square windows also showed surface Sn segregation, even though previous XRD ω-2θ peaks did not shift towards higher 2θ angle position. It suggested a non-uniformity in high Sn content GeSn SAG, leading to a mix of area of high Sn content/ no Sn droplets and area of low Sn content/ presence of Sn droplets, as previously observed on other CVD GeSn thin film growth [28,29]. Rectangle window, with higher fill factor, was clear of Sn segregation, agreeing well with earlier XRD ω-2θ data. In addition, unlike sample 2, EDX analysis on sample 5 showed extremely weak Sn incorporation in the 10 µm square window. In this case, a flat surface without {111} facets seemed to hint at a weak, or absent Sn incorporation in very small window sizes, in particular for window shapes with a low fill factor (circle, octagon, square).

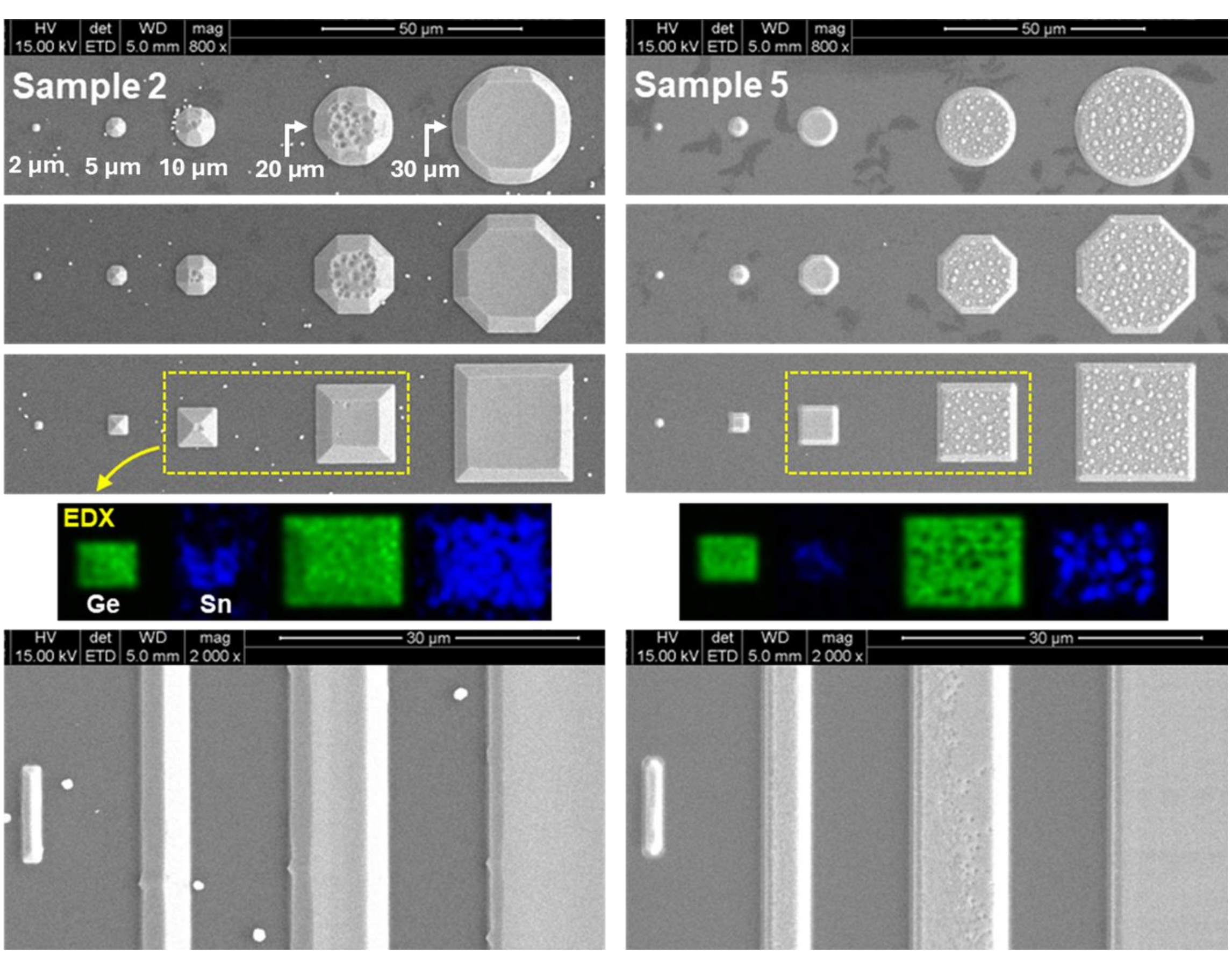

***Figure** 7. SEM images of GeSn SAG, for sample 2 (3.2% Sn - left) and sample 5 (8.7% Sn - right), on circle, octagon, square and rectangle window shape from 2 µm to 30 µm size. EDX characterization was performed on square windows of 10 µm and 20 µm size for both samples.*

For a further look into the impact of window size on Sn incorporation in GeSn SAG, micro-Raman characterization was conducted. We first analyzed the micro-Raman spectra on sample 4 - with relatively high Sn content of 7.1% Sn and without Sn segregation – on rectangle and square windows of 50 µm, 10 µm, 2 µm (**Figures 8a-f**). On 50 µm and 10 µm windows, both window shapes showed a red-shift of Ge($F_{2g}$) mode, as well as the presence of additional modes around 260 $cm^{-1}$ originated from Ge-Sn bond, indicating successful Sn incorporation [30,31]. At 2 µm windows, Ge($F_{2g}$) mode position of sample 4 closely matches that of Ge SAG (sample 0). For rectangle windows, Ge-Sn mode can still be identified, indicating successful Sn incorporation down to the smallest window, but possibly with reduced Sn content. For square windows, Ge-Sn mode disappeared, indicating an absence of Sn incorporation in this case.

In addition, we plotted the Ge($F_{2g}$) Raman shift in **Figures 8g-j** for all window shapes and sizes in sample 0 (Ge SAG), sample 2 (3.2% Sn), and sample 4 (7.1% Sn). Both GeSn SAG samples exhibited overall a red-shift of Ge($F_{2g}$) mode for window sizes below 40 µm, before a rapid re-increase between 10 µm and 2 µm. This red-shift of Ge($F_{2g}$) mode can be correlated with the appearance of {111} facets observed earlier from SEM images in **Figure 7**, which indicated an increase of relaxation degree in the GeSn active layer. At very small window sizes, below 10 µm, we attributed the resurgence of the Ge($F_{2g}$) peak position to lower or even absent Sn incorporation in the SAG island, as observed from **Figures 8c,f**. For rectangle shapes, there was a lesser resurgence of the Ge($F_{2g}$) mode compared to other shapes and it can still be clearly distinguished from that of Ge SAG, indicating indeed a successful, albeit lower Sn incorporation.

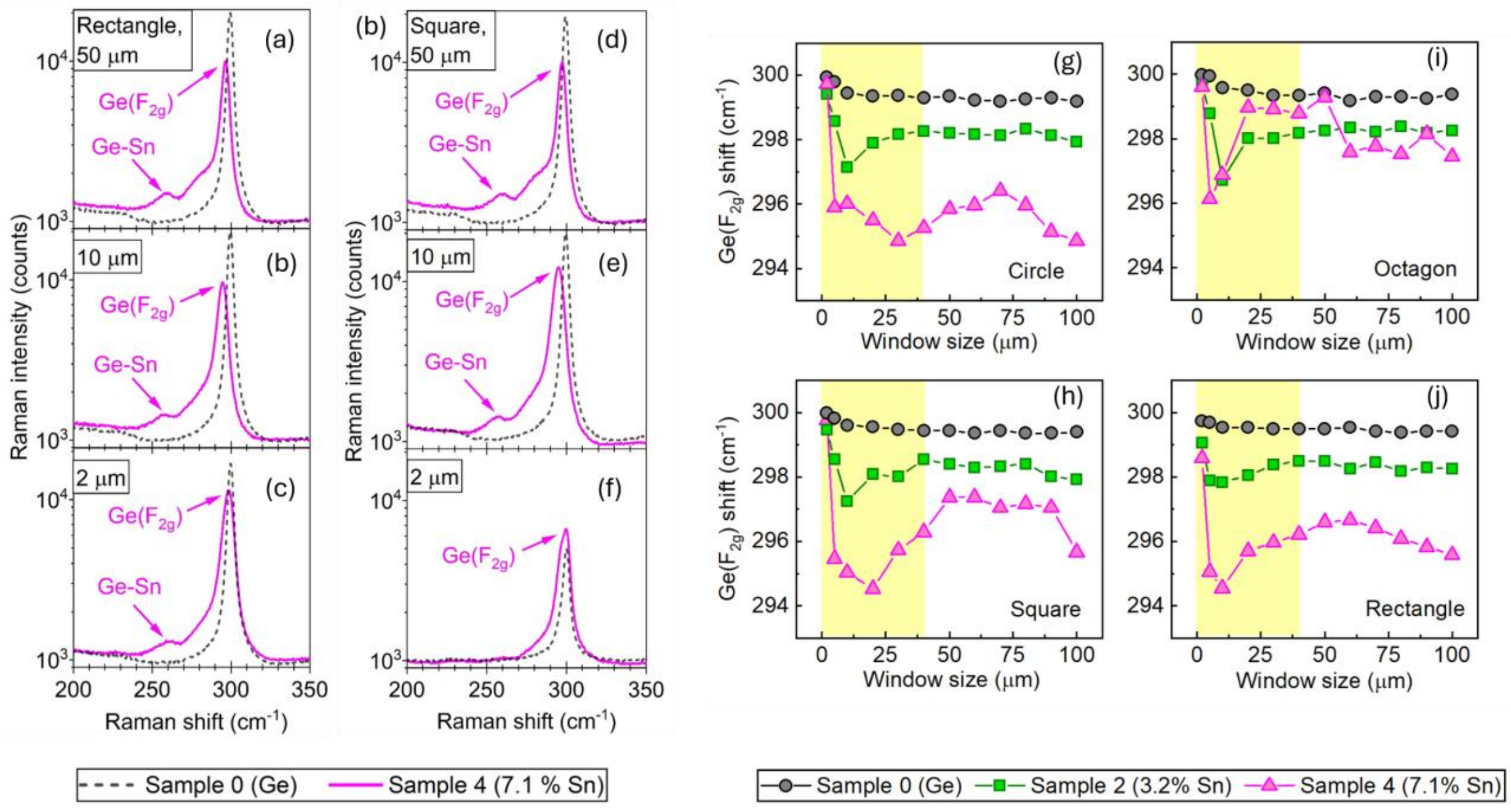

***Figure 8****. (a)-(c): Micro-Raman spectra of sample 4 (7.1% Sn) for rectangle windows of 2 µm, 10 µm and 50 µm width respectively. (d)-(f) Micro-Raman spectra of sample 4 (7.1% Sn) for square windows of 2 µm, 10 µm and 50 µm width respectively. Red arrows indicated the position of Ge-Sn and Ge($F_{2g}$) mode. (g)-(j) Ge($F_{2g}$) mode position in sample 0 (Ge), sample 2 (3.2% Sn) and sample 4 (7.1% Sn) as function of window size, for circle, square, octagon and rectangle window shapes respectively. Overall, window sizes below 40 µm – marked with yellow background – exhibited a significant fluctuation in the Ge($F_{2g}$) mode position, signaling a deviation between SAG and thin film growth characteristics.*

A thicker GeSn/Ge layer than the nominal target might be an explanation for the complex evolution of SAG growth characteristics below 40 µm of window size observed from micro-Raman data, correlated to an increase of relaxation degree and morphology change. Previous works of Ge/ SiGe SAG on Si [32–34] and III-V alloys (GaAs, InP) SAG on Si [35–37] reported an increase of growth rate as window size or fill factor shrank – known as loading effect - compared to thin film growth. It should be noted that thin film GeSn growth rate strongly depended on Sn content, with lower growth rate observed when Sn content increases due to reduced growth temperature [38–43]. It also depended on the choice of Ge precursor, with lower growth rate observed when using low reactivity precursors like $GeH_4$ in this work [39,42]. TEM characterization was thus conducted to analyze GeSn and Ge layer thickness from SAG (**Figure 9**). On square windows of 40 µm size (**Figures 9a,b**), we observed a clear overgrowth of both Ge buffer (810 nm thick) and GeSn layer (500 - 700 nm thick), deviated from our initial goal of confined growth inside $SiO_2$ window and thus provided evidence of loading effect in GeSn SAG. Overgrowth was observed to be more pronounced near the window edge on sample 2. On sample 5, one can observe the presence of Sn droplets both on the surface and in the bulk of grown GeSn layer, alongside degraded crystal quality. For high Sn content SAG recipe, forced high growth rate might create a different GeSn growth condition compared to thin film growth, the latter with expected low growth rate, leading to unsuccessful high Sn incorporation and presence of Sn droplets in the grown layer.

TEM characterization on rectangle SAG stripe of different width - 10 µm and 40 µm - further revealed how loading effect depended on window size (**Figures 9c,d**): increase of GeSn layer thickness from 220 nm to 530 nm with decreased stripe width showed that loading effect became more pronounced as window size shrank. A very strong loading effect at sub-10 µm window size might substantially change GeSn growth condition compared to thin film case to the point that Sn incorporation was no longer favored, as observed earlier from EDX and micro-Raman analysis. It should be noted that other factors, like Ge, Sn adatoms diffusion length can also have significant impact on very small micro-sized or nano-sized GeSn SAG, where the window size and adatoms diffusion length became comparable [44]. In addition, the choice of gas flow direction - horizontal flow to the wafer surface for the reactor used in this work, or perpendicular flow to the wafer surface (showerhead injection) in other reactor [38,45] – can also induce different growth kinetics on patterned substrate with deeply etched windows, which can lead to different behavior of loading effect and thus different SAG optimization strategies moving forward.

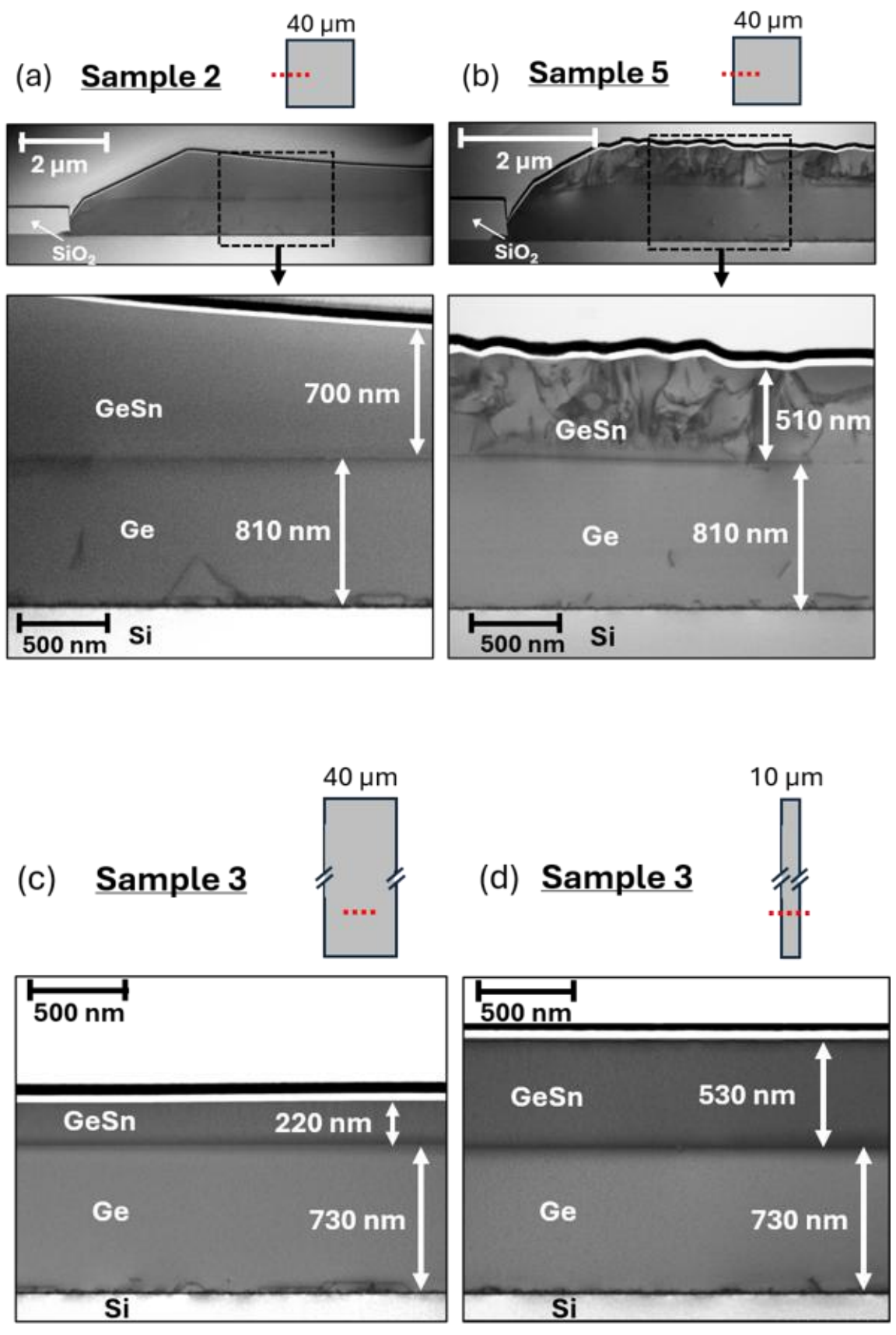

***Figure 9***. *(a), (b) TEM images of square windows of 40 µm size, from sample 2 (3.2% Sn) and sample 5 (8.7% Sn) respectively. Top images showed a zoomed-out view of the SAG island, with overgrowth of GeSn/Ge layers from the* $SiO_2$ *window (570 nm thick* $SiO_2$ *layer), while the bottom image showed a zoomed-in view, for area indicated in the black dashed windows. (c), (d) TEM*

*images of rectangle windows from sample 3, of 40 µm width and 10 µm width respectively. Red dashed lines indicated the TEM cut section positions – near the window edge.*

## CONCLUSION

In this work, we reported successful GeSn SAG with Sn content from 3.2% to 8.7% of good optical quality, demonstrated through tunable photoluminescence spectra and GeSn SAG photoconductor operation with detection range up to 2 µm of wavelength. Detailed analysis into SAG material characteristics revealed an increase of growth rate compared to thin film growth for SAG window size below 40 µm and in particular for low fill factor shapes. It either degraded crystal quality of SAG sample in high Sn content recipe with appearance of Sn droplets, or reduced and even inhibited Sn incorporation in very small window opening below 10 µm of size. Future work will focus on adjusting the growth recipe to recover thin film growth properties for small size SAG and high Sn content SAG, and on doping for GeSn SAG electrically injected laser and photodiode detector.

## METHODS

### Material growth and characterization

Ge and GeSn SAG samples, as well as GeSn thin film samples were epitaxially grown on a ASM Epsilon® 2000 RPCVD reactor, with $GeH_4$ and $SnCl_4$ as Ge and Sn gas precursors. XRD (004) ω-2θ and (224) RSM measurements were conducted using the Rigaku SmartLab Multipurpose X-ray Diffraction System equipped with a 2.0 kW Cu X-ray tube, four-bounce Ge(220) monochromator, and the HyPix-3000 pixel array multi-dimensional detector. PL measurement was conducted under 1064 nm laser excitation, using a Horiba iHR320 spectrometer equipped with an InSb detector. SEM characterization was performed with the FEI Nova Nanolab 200 system, using an accelerating voltage of 15 kV. Micro-Raman measurements were conducted using a Horiba Jobin-Yvon LabRAM HR microscope system equipped with 632.8 nm laser excitation, an Andor Si CCD detector and a 100× objective lens, resulting in focused laser spot of 2 μm diameter.

**Photoconductor fabrication and characterization**

For photoconductor fabrication, standard lithography was used to define window opening pattern on $SiO_2$ layer and gold deposition pattern on SAG photoconductor. Ohmic contact was placed on SAG island using 10/300 nm Cr/Au e-beam deposition. Photoconductor characterization was performed from 77 K to 300 K, with I-V and spectral response measurement. I-V characteristics were measured with Keysight B2911B source measure unit (SMU). 1.55 μm responsivity were measured using 1.55 μm laser source from BKTel Photonics. The incoming optical power was measured using Thorlabs S401C thermal sensor. Responsivities were calculated either from directly extracting dark I-V from light I-V. Finally, spectral responses were measured using a ThermoFisher Nicolet 8700 FTIR, equipped with an IR source. The raw FTIR spectra were then converted to responsivity spectra using calibrated data from a commercial InGaAs detector with detection cutoff wavelength of 2.6 μm.

## AUTHOR INFORMATION

**Corresponding Author**

* Shui-Qing Yu

Email: syu@uark.edu

**Author Contributions**

S.-Q. Y., W. D., G. S, B. L. and J. L. proposed and supervised the projects. P. C. G. designed the window mask used for SAG process. M. C. and S. M. produced $SiO_2$ patterned wafers for SAG. F. M. O. and H. S. conducted micro-Raman measurements and analysis. K. D., X. M. and S. A. conducted PL measurement of thin film samples and SAG samples. H. S. conducted XRD measurement and analysis. H. S. and A. G. conducted SEM-EDX characterization. R. K. fabricated SAG photoconductor devices. Q. M. T. and J. R. conducted SAG photoconductor characterization and data analysis. H. S. and Q. M. T. performed TEM data analysis. All authors discussed the results and commented on the manuscript.

‡ H. S. and Q. M. T. contributed equally to this work.

**ACKNOWLEDGMENT**

This work is supported by United States Air Force under Contract No. FA865023C1140, the Department of Defense Small Business Technology Transfer (STTR) Program, Office of the Secretary of Defense-Basic Research Office (OSD-BRO) and the Army Research Office under Contract No. W911NF-24-C-0052, the Air Force Research Laboratory (AFRL/RYDHS and AFRL/RYKSE) under Contract No. FA237724CB033, and Office of Naval Research under Grant No. N000142412651.

Lincoln Laboratory at Massachusetts Institute of Technology would like to acknowledge the support as follows: *DISTRIBUTION STATEMENT A. Approved for public release. Distribution is unlimited. This material is based upon work supported by the Department of the Air Force under Air Force Contract No. FA8702-15-D-0001 or FA8702-25-D-B002. © 2026 Massachusetts Institute of Technology. Delivered to the U.S. Government with Unlimited

Rights, as defined in DFARS Part 252.227-7013 or 7014 (Feb 2014). Notwithstanding any copyright notice, U.S. Government rights in this work are defined by DFARS 252.227-7013 or DFARS 252.227-7014 as detailed above. Use of this work other than as specifically authorized by the U.S. Government may violate any copyrights that exist in this work. Any opinions, findings, conclusions or recommendations expressed in this material are those of the author(s) and do not necessarily reflect the views of the Department of the Air Force.

We would also like to acknowledge Dr. Yuriy Mazur for his assistance on micro-Raman characterization, and Dr. Pargam Vashishtha for his support on SEM-EDX measurement, and Abdulla Said Ali for his support on SAG photoconductor measurement.

**REFERENCES**

(1) Wirths, S.; Geiger, R.; von den Driesch, N.; Mussler, G.; Stoica, T.; Mantl, S.; Ikonic, Z.; Luysberg, M.; Chiussi, S.; Hartmann, J.-M.; Sigg, H.; Faist, J.; Buca, D.; Grutzmacher, D. Lasing in Direct-Bandgap GeSn Alloy Grown on Si. *Nat. Photonics* 2015, *9*, 88.

(2) Chretien, J.; Thai, Q. M.; Frauenrath, M.; Casiez, L.; Chelnokov, A.; Reboud, V.; Hartmann, J.-M.; El-Kurdi, M.; Pauc, N.; Calvo, V. Room Temperature Optically Pumped GeSn Microdisk Lasers. *Appl. Phys. Lett.* 2022, *120*, 051107.

(3) Bjelajac, A.; Gromovyi, M.; Sakat, E.; Wang, B.; Patriarche, G.; Pauc, N.; Calvo, V.; Boucaud, P.; Boeuf, F.; Chelnokov, A.; Reboud, V.; Frauenrath, M.; Hartmann, J.-M.; El-Kurdi, M. Up to 300 K Lasing with GeSn-On-Insulatormicrodisk Resonators. *Opt. Express* 2022, *30* (3), 3954.

(4) Buca, D.; Bjelajac, A.; Spirito, D.; Concepción, O.; Gromovyi, M.; Sakat, E.; Lafosse, X.; Ferlazzo, L.; von den Driesch, N.; Ikonic, Z.; Grützmacher, D.; Capellini, G.; El Kurdi, M. Room Temperature Lasing in GeSn Microdisks Enabled by Strain Engineering. *Adv. Opt. Mater.* 2022, *10* (22), 2201024.

(5) Zhou, Y.; Miao, Y.; Ojo, S.; Tran, H.; Abernathy, G.; Grant, J. M.; Amoah, S.; Salamo, G.; Du, W.; Liu, J.; Margetis, J.; Tolle, J.; Zhang, Y.; Sun, G.; Soref, R. A.; Li, B.; Yu, S.-Q. Electrically Injected GeSn Lasers on Si Operating up to 100 K. *Optica* 2020, *7* (8), 924–928.

(6) Zhou, Y.; Ojo, S.; Wu, C.-W.; Miao, Y.; Tran, H.; Grant, J. M.; Abernathy, G.; Amoah, S.; Bass, J.; Salamo, G.; Du, W.; Chang, G.-E.; Liu, J.; Margetis, J.; Tolle, J.; Zhang, Y.-

H.; Sun, G.; Soref, R. A.; Li, B.; Yu, S.-Q. Electrically Injected GeSn Lasers with Peak Wavelength up to 2.7 Mm. *Photonics Res.* 2022, *10* (1), 222–229.

(7) Acharya, S.; Stanchu, H.; Kumar, R.; Ojo, S.; Alher, M.; Benamara, M.; Chang, G.-E.; Li, B.; Du, W.; Yu, S.-Q. Electrically Injected Mid-Infrared GeSn Laser on Si Operating at 140 K. *IEEE Journal of Selected Topics in Quantum Electronics* 2025, *31* (1: SiGeSn Infrared Photon. and Quantum Electronics), 1–7.

(8) Tran, H.; Pham, T.; Margetis, J.; Zhou, Y.; Dou, W.; Grant, P. C.; Grant, J. M.; Al-Kabi, S.; Sun, G.; Soref, R. A.; Tolle, J.; Zhang, Y.-H.; Du, W.; Li, B.; Mortazavi, M.; Yu, S.-Q. Si-Based GeSn Photodetectors toward Mid-Infrared Imaging Applications. *ACS Photonics* 2019, *6* (11), 2807–2815.

(9) Talamas Simola, E.; Kiyek, V.; Ballabio, A.; Schlykow, V.; Frigerio, J.; Zucchetti, C.; De Iacovo, A.; Colace, L.; Yamamoto, Y.; Capellini, G.; Grützmacher, D.; Buca, D.; Isella, G. CMOS-Compatible Bias-Tunable Dual-Band Detector Based on GeSn/Ge/Si Coupled Photodiodes. *ACS Photonics* 2021, *8* (7), 2166–2173.

(10) Thai, Q. M.; Kumar, R.; Ali, A. S.; Rudie, J.; Akwabli, S.; Qiu, Y.; Benamara, M.; Stanchu, H.; Dahal, K.; Ma, X.; Acharya, S.; Chang, C.-C.; Forcherio, G. T.; Claflin, B.; Du, W.; Yu, S.-Q. Systematic Study of High Performance GeSn Photodiodes with Thick Absorber for SWIR and Extended SWIR Detection. *arXiv:2602.15223* 2026.

(11) Waldron, N.; Wang, G.; Nguyen, N. D.; Orzali, T.; Merckling, C.; Brammertz, G.; Ong, P.; Winderickx, G.; Hellings, G.; Eneman, G.; Caymax, M.; Meuris, M.; Horiguchi, N.; Thean, A. Integration of InGaAs Channel N-MOS Devices on 200mm Si Wafers Using the Aspect-Ratio-Trapping Technique. *ECS Trans.* 2012, *45* (4), 115.

(12) De Koninck, Y.; Caer, C.; Yudistira, D.; Baryshnikova, M.; Sar, H.; Hsieh, P.-Y.; Özdemir, C. I.; Patra, S. K.; Kuznetsova, N.; Colucci, D.; Milenin, A.; Yimam, A. A.; Morthier, G.; Van Thourhout, D.; Verheyen, P.; Pantouvaki, M.; Kunert, B.; Van Campenhout, J. GaAs Nano-Ridge Laser Diodes Fully Fabricated in a 300-Mm CMOS Pilot Line. *Nature* 2025, *637* (8044), 63–69.

(13) Park, J.-S.; Bai, J.; Curtin, M.; Adekore, B.; Carroll, M.; Lochtefeld, A. Defect Reduction of Selective Ge Epitaxy in Trenches on Si(001) Substrates Using Aspect Ratio Trapping. *Appl. Phys. Lett.* 2007, *90* (5), 052113.

(14) Ackland, B.; Rafferty, C.; King, C.; Aberg, I.; O'Neill, J.; Sriram, T.; Lattes, A.; Godek, C.; Pappas, S. A Monolithic Ge-on-Si CMOS Imager for Short Wave Infrared. In *International Image Sensors Workshop*; International Image Sensors Society, 2009.

(15) Wirths, S.; Mayer, B. F.; Schmid, H.; Sousa, M.; Gooth, J.; Riel, H.; Moselund, K. E. Room-Temperature Lasing from Monolithically Integrated GaAs Microdisks on Silicon. *ACS Nano* 2018, *12* (3), 2169–2175.

(16) Staudinger, P.; Moselund, K. E.; Schmid, H. Exploring the Size Limitations of Wurtzite III–V Film Growth. *Nano Lett.* 2020, *20* (1), 686–693.

(17) Schmid, H.; Borg, M.; Moselund, K.; Gignac, L.; Breslin, C. M.; Bruley, J.; Cutaia, D.; Riel, H. Template-Assisted Selective Epitaxy of III–V Nanoscale Devices for Co-Planar Heterogeneous Integration with Si. *Appl. Phys. Lett.* 2015, *106* (23), 233101.

(18) Stanchu, H.; Abernathy, G.; Grant, J.; de Oliveira, F. M.; Mazur, Y. I.; Liu, J.; Du, W.; Li, B.; Salamo, G. J.; Yu, S.-Q. Development of Aspect Ratio Trapping Growth of GeSn on Si for Midwave Infrared Applications. *Journal of Vacuum Science & Technology B* 2024, *42* (4), 042802.

(19) Dou, W.; Benamara, M.; Mosleh, A.; Margetis, J.; Grant, P.; Zhou, Y.; Al-Kabi, S.; Du, W.; Tolle, J.; Li, B.; Mortazavi, M.; Yu, S.-Q. Investigation of GeSn Strain Relaxation and Spontaneous Composition Gradient for Low-Defect and High-Sn Alloy Growth. *Sci. Rep.* 2018, *8* (1), 5640.

(20) Assali, S.; Nicolas, J.; Moutanabbir, O. Enhanced Sn Incorporation in GeSn Epitaxial Semiconductors via Strain Relaxation. *J. Appl. Phys.* 2019, *125* (2), 025304.

(21) Stanchu, H. V; Kuchuk, A. V; Mazur, Y. I.; Margetis, J.; Tolle, J.; Yu, S.-Q.; Salamo, G. J. Strain Suppressed Sn Incorporation in GeSn Epitaxially Grown on Ge/Si(001) Substrate. *Appl. Phys. Lett.* 2020, *116* (23), 232101.

(22) Stanchu, H.; Kryvyi, S.; Margiotta, S.; Cook, M.; Grant, J.; Tran, H.; Acharya, S.; de Oliveira, F. M.; Mazur, Y. I.; Benamara, M.; King, C. A.; Du, W.; Li, B.; Salamo, G.; Yu, S.-Q. Comprehensive Material Study of Ge Grown by Aspect Ratio Trapping on Si Substrate. *J. Phys. D Appl. Phys.* 2024, *57* (25), 255107.

(23) Conley, B. R.; Mosleh, A.; Ghetmiri, S. A.; Du, W.; Soref, R. A.; Sun, G.; Margetis, J.; Tolle, J.; Naseem, H. A.; Yu, S.-Q. Temperature Dependent Spectral Response and Detectivity of GeSn Photoconductors on Silicon for Short Wave Infrared Detection. *Opt. Express* 2014, *22* (13), 15639–15652.

(24) Conley, B. R.; Margetis, J.; Du, W.; Tran, H.; Mosleh, A.; Ghetmiri, S. A.; Tolle, J.; Sun, G.; Soref, R.; Li, B.; Naseem, H. A.; Yu, S.-Q. Si Based GeSn Photoconductors with a 1.63 A/W Peak Responsivity and a 2.4 Mm Long-Wavelength Cutoff. *Appl. Phys. Lett.* 2014, *105* (22), 221117.

(25) Vines, P.; Kuzmenko, K.; Kirdoda, J.; Dumas, D. C. S.; Mirza, M. M.; Millar, R. W.; Paul, D. J.; Buller, G. S. High Performance Planar Germanium-on-Silicon Single-Photon Avalanche Diode Detectors. *Nat. Commun.* 2019, *10* (1), 1086.

(26) Rogalski, A. *Infrared and Terahertz Detectors*; CRC Press, 2022.

(27) Tran, R.; Xu, Z.; Radhakrishnan, B.; Winston, D.; Sun, W.; Persson, K. A.; Ong, S. P. Surface Energies of Elemental Crystals. *Sci. Data* 2016, *3* (1), 160080.

(28) Aubin, J.; Hartmann, J.-M.; Gassenq, A.; Rouviere, J. L.; Robin, E.; Delaye, V.; Cooper, D.; Mollard, N.; Reboud, V.; Calvo, V. Growth and Structural Properties of Step-Graded, High Sn Content GeSn Layers on Ge. *Semicond. Sci. Technol.* 2017, *32* (9).

(29) Rosson, N.; Acharya, S.; Fischer, A. M.; Collier, B.; Ali, A.; Torabi, A.; Du, W.; Yu, S.-Q.; Scott, R. C. Development of GeSn Epitaxial Films with Strong Direct Bandgap Luminescence in the Mid-Wave Infrared Region Using a Commercial Chemical Vapor Deposition Reactor. *Journal of Vacuum Science & Technology B* 2024, *42* (5), 052210.

(30) D'Costa, V. R.; Tolle, J.; Roucka, R.; Poweleit, C. D.; Kouvetakis, J.; Menéndez, J. Raman Scattering in Ge1−ySny Alloys. *Solid State Commun.* 2007, *144* (5), 240–244.

(31) Bouthillier, É.; Assali, S.; Nicolas, J.; Moutanabbir, O. Decoupling the Effects of Composition and Strain on the Vibrational Modes of GeSn Semiconductors. *Semicond. Sci. Technol.* 2020, *35* (9), 095006.

(32) Ishitani, A.; Endo, N.; Tsuya, H. Local Loading Effect in Selective Silicon Epitaxy. *Jpn. J. Appl. Phys.* 1984, *23* (6A), L391.

(33) Menon, C.; Bentzen, A.; Radamson, H. H. Loading Effect in SiGe Layers Grown by Dichlorosilane- and Silane-Based Epitaxy. *J. Appl. Phys.* 2001, *90* (9), 4805–4809.

(34) Bodnar, S.; Morin, C.; Regolini, J. L. Single-Wafer Si and SiGe Processes for Advanced ULSI Technologies. *Thin Solid Films* 1997, *294* (1), 11–14.

(35) Olsson, F.; Zhu, T.; Mion, G.; Lourdudoss, S. Large Mask Area Effects in Selective Area Growth. *J. Cryst. Growth* 2006, *289* (1), 24–30.

(36) Brammertz, G.; Mols, Y.; Degroote, S.; Leys, M.; Van Steenbergen, J.; Borghs, G.; Caymax, M. Selective Epitaxial Growth of GaAs on Ge by MOCVD. *J. Cryst. Growth* 2006, *297* (1), 204–210.

(37) Dupuis, N.; Décobert, J.; Lagrée, P.-Y.; Lagay, N.; Carpentier, D.; Alexandre, F. Demonstration of Planar Thick InP Layers by Selective MOVPE. *J. Cryst. Growth* 2008, *310* (23), 4795–4798.

(38) Concepción, O.; Søgaard, N. B.; Bae, J.-H.; Yamamoto, Y.; Tiedemann, A. T.; Ikonic, Z.; Capellini, G.; Zhao, Q.-T.; Grützmacher, D.; Buca, D. Isothermal Heteroepitaxy of Ge1–XSnx Structures for Electronic and Photonic Applications. *ACS Appl. Electron. Mater.* 2023, *5* (4), 2268–2275.

(39) Hartmann, J. M.; Marion, T. Impact of Flows, Temperature and Pressure on the GeSn Growth Kinetics with a Digermane and Tin Tetrachloride Chemistry. *Mater. Sci. Semicond. Process.* 2024, *169*, 107893.

(40) Khazaka, R.; Aubin, J.; Nolot, E.; Hartmann, J.-M. Investigation of the Growth of Si-Ge-Sn Pseudomorphic Layers on 200 Mm Ge Virtual Substrates: Impact of Growth Pressure, HCl and Si2H6 Flows. *ECS Trans.* 2018, *86* (7), 207.

(41) Jahandar, P.; Weisshaupt, D.; Colston, G.; Allred, P.; Schulze, J.; Myronov, M. The Effect of Ge Precursor on the Heteroepitaxy of Ge1−xSnx Epilayers on a Si (001) Substrate. *Semicond. Sci. Technol.* 2018, *33* (3), 034003.

(42) Hartmann, J. M.; Marion, T. An Assessment of Germane and Tin Tetrachloride for GeSn Epitaxy. *J. Cryst. Growth* 2025, *667*, 128280.

(43) Hartmann, J.-M.; Frauenrath, M.; Richy, J. Epitaxy of Pseudomorphic GeSn Layers with Germane (GeH4) or Digermane (Ge2H6) as Ge Precursors and Tin Tetrachloride (SnCl4) as the Sn Precursor. *ECS Trans.* 2020, *98* (5), 225.

(44) Li, Q.; Krauss, J. L.; Hersee, S.; Han, S. M. Probing Interactions of Ge with Chemical and Thermal SiO2 to Understand Selective Growth of Ge on Si during Molecular Beam Epitaxy. *The Journal of Physical Chemistry C* 2007, *111* (2), 779–786.

(45) Wirths, S.; Buca, D.; Tiedemann, A. T.; Holländer, B.; Bernardy, P.; Stoica, T.; Grützmacher, D.; Mantl, S. Epitaxial Growth of Ge1-XSnx by Reduced Pressure CVD Using SnCl4 and Ge2H6. *ECS Trans.* 2013, *50* (9), 885.